\documentclass[12pt]{article}

\AtBeginDocument{
	\addtocontents{toc}{\normalsize}%\footnotesize
}
\pdfoutput=1

\usepackage{amsmath,amssymb,amsfonts,amsthm,amscd,mathrsfs}
\usepackage{esvect}
\usepackage{xcolor}
\definecolor{darkblue}{rgb}{0.1,0.1,.7}
\usepackage[]{version}
\usepackage[]{graphicx}
\usepackage[]{latexsym}
\usepackage{geometry}
\usepackage[all,cmtip]{xy}

\usepackage[margin=10pt,font=small,labelfont=bf]{caption}
\usepackage{ifthen}
\usepackage{soul}
\usepackage{tikz}
\usepackage{array,mathrsfs,amsfonts,yfonts,dsfont,bbm,colonequals}
\usepackage[nodisplayskipstretch]{setspace}
\usepackage{dsfont}
\usepackage{cite}
\usepackage{xspace}
\usepackage{empheq}
\usepackage{extarrows}
\usepackage{braket}

\usepackage{xcolor}
\usepackage[colorlinks, linkcolor=darkblue, citecolor=darkblue, urlcolor=darkblue, linktocpage]{hyperref} 

\usepackage{subcaption}
\usepackage[utf8]{inputenc}
\usepackage{setspace}
\usepackage{float, graphicx}
\usepackage{mathrsfs}

\def\mc{\mathcal}

\numberwithin{equation}{section}

\begin{document}
	
	\vspace*{-.6in} \thispagestyle{empty}
	\begin{flushright}

	\end{flushright}
	%%
	%% Title
	%%
	\vspace{1cm} {\Large
		\begin{center}
			{\bf On multipoint Ward identities for\\ superconformal line defects}\\
	\end{center}}
	\vspace{1cm}
	%%%
	%% Authors
	%%%
	\begin{center}
		{\bf Gabriel Bliard${}^a$}\\[1cm] 
		{
			\small
			{\em ${}^a$Laboratoire de Physique, , \\ \'Ecole Normale Sup\'erieure, \\
   Universit{\'e} PSL, CNRS, Sorbonne Universit{\'e}, Universit{\'e} Paris Cit{\'e}, \\
24 rue Lhomond, F-75005 Paris, France}

			\normalsize
		}
		
		%\vspace{1cm}\today
	\end{center}
	
	\begin{center}
		{  \texttt{\  gabriel.bliard@ens.fr} 
		}
		\\
		%\vspace{1cm}\today
	\end{center}
	
	\vspace{8mm}
	
	\begin{abstract}
 \vspace{2mm}
%The constraints from the Supersymmetry leading to the s
Superconformal Ward identities are revisited in the context of superconformal line defects. 
Multipoint correlators of topological operators inserted on superconformal lines are studied. In particular, it is known that protected operators preserving enough of the supersymmetry become topological after performing a topological twist. By definition, such a correlator is constant in the topological limit. By analysing the topological constraint on the OPE of such operators, the correlator is further constrained away from this limit. The constraints on multipoint correlators match the known superconformal Ward identities in the case of 4-point functions. This allows for an simple and universal derivation of the superconformal Ward identities governing the multipoint correlation functions of such operators. This concept is illustrated by 1/2-BPS operators with an $su(2)$ R-symmetry and further explored in the case of the displacement multiplet on the 1/2-BPS Wilson line in 4d $\mathcal{N}=4$ super Yang-Mills theory supporting the conjectured multipoint Ward identities in the literature.

	\end{abstract}
	\vspace{2in}

	\newpage
	
	{
		\setlength{\parskip}{0.05in}
		\tableofcontents
		\renewcommand{\baselinestretch}{1.0}\normalsize
	}
	
	%\newpage
	
	\setlength{\parskip}{0.1in}
 \setlength{\abovedisplayskip}{15pt}
 \setlength{\belowdisplayskip}{15pt}
 \setlength{\belowdisplayshortskip}{15pt}
 \setlength{\abovedisplayshortskip}{15pt}
 
	% \newpage
 \section{Introduction}
Superconformal field theories (SCFTs) play a central role in modern theoretical physics, appearing both in the context of string theory through the AdS/CFT correspondence and as supersymmetric extensions to critical phenomena and condensed matter systems. While the additional symmetries makes the systems less rich in complexity, there is still much to be known and we are far from a complete description of these theories. On the other hand, these symmetries allow us to delve into more complex problems. One such problem is that of extended operators such as conformal defects. These arise naturally in any CFT \cite{Billo:2016cpy} and their understanding is essential to a full description of the theory. In fact, part of the additional data which is introduced by these non-local operators can be studied as an independent defect-SCFT where the broken symmetries of the embedding theory play a special role as operator insertions. In this context, much work has been done, mainly on four-point functions. \par 
Superconformal theories are extremely constrained; on one hand, the conformal theory reduces the study of all correlators to a countable set, the conformal data $\{\Delta,c_{123}\}$, on the other, the supersymmetry protects the dimension of special operators and provides tools such as localisation \cite{Giombi:2018qox} to evaluate some of the conformal data at finite coupling. Such operators preserving half the supersymmetries -1/2 BPS- generically depend on the position of insertion and their polarisation under the R-symmetry. The constraints coming from the supersymmetry for four-point correlators of such operators in \cite{Dolan:2004mu, Eden:2000qp, Nirschl:2004pa,Hartwell:1994rp} leads to relations between the position and the polarisation cross-ratios; the superconformal Ward identities (SCWI).\par 
These SCWI constrain the form of correlators in the theory and as such are a very useful tool which reduces the number of variables and the difficulty in the study of a 4-point correlator. This has been used in \cite{Dolan:2004iy,Beem:2013qxa,Alday:2014qfa,Chester:2014fya,Beem:2014zpa,Belitsky:2014zha} in this context to study the four-point correlators with more ease. The recent interest in multipoint correlators justifies the search for more general Ward identities. For example, in \cite{Giombi:2023zte}, the superconformal Ward identities could be used to extract the CFT data from the single R-symmetry channel computed via the conformal gauge. The resulting understanding can help extend the numerical bootstrap of 6-point correlators\cite{Antunes:2023dlk} from that of scalars to that of constrained supermultiplets. The R-symmetry dependence of correlators is then reduced and the space-time dependent part can then effectively be studied using other methods such as Witten diagrams \cite{Freedman:1998tz,Rastelli:2017udc,Bliard:2022xsm,Paulos:2011ie} and Mellin amplitudes\cite{Mack:2009mi,Penedones:2010ue}. \par
In the context of superconformal line defect correlators, the case is even clearer as there is only one spacetime cross-ratio for four external insertions. In this case, the SCWIs reduce the problem of a three-variable function to that of a single function of a single cross-ratio \cite{Liendo:2016ymz}. This was then used to constrain the CFT data through the numerical bootstrap \cite{Liendo:2018ukf}, to fix the perturbative form of the correlator at weak \cite{Cavaglia:2022qpg,Cavaglia:2021bnz} and strong coupling \cite{Ferrero:2021bsb}. Being a devil's advocate, one could ask why bother with higher-point functions since the OPE relates these all to 3-point functions and can therefore be found by looking at the 4-point functions. However, just as 4-point correlators contain an infinite number of 3-point functions, 6-point functions contain an infinite number of 4-point functions and so on. On one hand, this means that certain sectors of the CFT data corresponding to operators which are not exchanged in the known  4-point functions can be accessed\cite{Antunes:2023dlk,Barrat:2024nod,Bliard:2023zpe}. Furthermore, there is a technical aspect wherein it is easier to look at a single higher-point function than to consider an infinite number of correlators. On the other hand, there is the technical difficulty of the mixing of correlators. While this is a feature of perturbations in CFTs so does not concern methods  such as the numerical bootstrap, it certainly is a roadblocks when considering the analytic bootstrap such as in \cite{Ferrero:2023gnu,Ferrero:2023znz}. In this case, resolving the mixing problem requires the authors to analyse correlators with generic long multiplets as external operators and requires a detailed construction of both the anomalous dimension matrix and the long operators themselves at strong coupling. The way that higher-point functions avoid such an issue is two-fold. First, the additional operators allow for the analysis of long operators through the OPE without the need for an explicit construction of an operator in terms of the elementary fields. Furthermore, when organising the exchanged operators in terms of their 'length', the free theory already contains non-trivial information. Finally, there is an upper limit to the highest-point function which contains all the relevant information in a theory\cite{Antunes:2023dlk}.\par 
On another hand, protected operators such as the displacement multiplet in supersymmetric theories play a special role in describing deformations of the line \cite{Cooke:2017qgm,Drukker:2006xg,Correa:2012at,Gaberdiel:2008fn}. In this context, understanding higher-point functions can help describe deformations and setups different from the usual cusp solution \cite{Correa:2012nk}.
As such, the analysis of multipoint correlators in the context of supersymmetric defect theories is essential in the understanding of the CFT data and the deformations of the defect. Despite the many symmetries at play, understanding these correlators away from the topological limit is challenging. The many R-symmetry channels transform the bosonic case of a function depending on $n-3$ variables to as many functions as there are channels. In the case of 4 external operators, these are related by the SCWI which reduces the problem to a single function of a single variable. This note explores the generalisation of this using a different method as that used in \cite{Dolan:2004iy,Liendo:2016ymz}. This method not only allows for the derivation of the SCWI in a highly adaptable manner, but also explains why the number of constraints from the SCWI increases as the number of insertions increases despite the symmetry group staying the same. Such a method is analogous to the deriving of superconformal blocks but uses the vanishing of the eigenvalue to constrain the full correlator instead of individual blocks.\par 
The note proceeds as follows; the first section \ref{Sec: Section 1} will be focused on the derivation of the SCWI for multipoint insertions with an R-symmetry group $SU(2)$. Since the R-symmetry and the conformal group act in a very similar way, this leads to the simplest multipoint Ward identities. The section includes the three key elements to the derivation, in the first, the concept of the topological sector and the topological twist such as is \cite{Drukker:2009sf} are reviewed and the relevant topologically twisted conformal algebra given explicitly. In the second, the operator product expansion between two topological operators is studied and the constraints from the quadratic Casimir of the twisted conformal algebra are shown. In the final section, the linearisation of the quadratic Casimir as a differential equation gives the generalisation of the superconformal Ward identities. 
\begin{align}
(\partial_{\chi_i}+\partial_{\rho_i})f(\chi_i,\rho_i)|_{\rho_i\rightarrow \chi_i}=0
\end{align}
where $\chi_i$ are the space-time cross ratios, $\rho_i$ are the R-symmetry cross ratios, and the topological limit $\rho_i\rightarrow\chi_i$ is taken.
In the second section, the multipoint correlators of 1/2-BPS insertions on the 1/2-BPS Wilson line defect in $\mc{N}=4$ super Yang-Mills (SYM) are considered. As expected the equivalent generalisation of the known SCWIs match the 4-point results and support the conjecture for multipoint Ward identities made in \cite{Barrat:2021tpn} in a certain limit.
\begin{align}
(\frac 12\partial_{\chi_{i^*}} +\partial_{\zeta_{i^*}})f(\chi_i,\zeta_i\bar{\zeta}_i,(1-\zeta_i)(1-\bar{\zeta}_i),(\zeta_i-\zeta_j)\bar{\eta_{ij}})\raisebox{-.2cm}{$\Biggr|$}_{\raisebox{+.2cm}{$\begin{subarray}{l}
				\zeta_i\rightarrow \chi_i \\\eta_{ij}\rightarrow (\chi_i-\chi_j)\\\bar{\zeta}_{i^*}\rightarrow 1
			\end{subarray}$}}=0.
\end{align}
where $\chi_i$ are the $n-3$ space-time cross-ratios, $\zeta_i,\bar{\zeta}_i,\eta_{ij}$ are the $2(n-3)+\frac 12 (n-3)(n-4)$ R-symmetry cross-ratios. These Ward identities explains why the number of constraints increases with the number of external insertions and agrees with the conjectured Ward identities in \cite{Barrat:2021tpn}.
A number of technical appendices complete this note. 

%\section{Notation}
%The superfields will be written as 
%\begin{align}
%    \Phi_{\Delta,\lambda}(t,\theta_a,\bar{\theta}^a,Y^a) =_{\theta\rightarrow 0} \phi(t,Y^a)
%\end{align}
%where $\phi$ is the superconformal primary as a function of position and R-symmetry polarisation. It has eigenvalues $\Delta,\lambda$ under the dilation operator and R-symmetry Cartan respectively. In general $\lambda$ is a vector. We will not need any of the description for the superconformal descendents. Correlators on the line will be written as 
%\begin{align}
%    \langle \phi(t_1) \phi(t_2)\phi(t_3)...\phi(t_n)\rangle  = 
%\end{align}
%\begin{itemize}
%\item Why is it relevant
%\item Where does it appear
%\item How does it help
%\item Current and future challenges
%\item Notation
%\end{itemize}

 \section{The simplest multipoint Ward identities}\label{Sec: Section 1}
 The case of an $\mc{N}=2$ supersymmetry with $su(2)$ R-symmetry is particularly enlightening because it will appear as a subgroup of all the other cases considered. In this case, the Ward identities were found over 20 years ago in \cite{Dolan:2004mu} and the superconformal blocs derived recently in \cite{Baume:2019aid}. In order to generalise these results to a larger number of insertions, the relevant Ward identities must be found. The method above consists of fixing the superconformal cross-ratios using the symmetries of the correlator. The variation under a generic supersymmetric transformation then generates terms with spurious poles whose vanishing denominator gives the Ward identities. This powerful formalism \cite{Dolan:2004mu} is universal for four-point correlators and has been used to derive many of the supersymmetric Ward identities present in the literature \cite{Liendo:2016ymz,Baume:2019aid}. The question then arises in how to extend this to higher-point correlators. The first problem arising is that the supersymmetry is no longer enough to fully fix the fermionic coordinates to 0. This in general leads to nilpotent (or fermionic) cross-ratios. The resulting independent functions can then be fixed by the OPE and considering the superconformal Casimir such as in \cite{Cornagliotto:2017dup,Buric:2020zlb,Bissi:2015qoa,Buric:2019rms,Gimenez-Grau:2019hez}. Given these additional terms, it is unclear whether the same amount of supersymmetry will generate more constraints for an increasing number of external operators. The results from this note suggests that this is indeed the case and provides an intuitive physical understanding to why that is the case. In thius sense, one would expect that extending the work of \cite{Dolan:2004mu} to higher-point functions will indeed lead to many spurious poles in the coincident limit of space-time and R-symmetry variables and correspondingly, many superconformal Ward identities. \par 
The approach this note takes compared to the method described above is rather a detour on the superconformal constraints. The goal is to develop a set of purely bosonic identities satisfied by the correlator of the superconformal primary of the multiplet. The supersymmetry then works in the background. The main advantage of this formalism is purely technical: Once the system is set up, one can forget entirely about supersymmetry and consider only bosonic quantities which greatly reduces the difficulty of the problem. As such, these Ward identities have been derived for up to 12 external operators. While this may seem reductive, many of the results and much of the CFT data in superconformal defect setups \cite{Bianchi:2020hsz,Liendo:2018ukf,Barrat:2021yvp} can be obtained by considering purely bosonic functions. \par  The concept
will be illustrated in the context of $su(2)$ R-symmetry and 1d conformal defects. However, this only relies on a $Q-$exact conformal algebra which also exists in other setups \cite{Drukker:2008pi,Meneghelli:2022gps}. The first subsection will establish the topological sector and the twisted conformal algebra which is $Q$-exact under supercharges which annihilate the superprimary. One can then act on the operators with the twisted translations therefore annihilating the correlator; it no longer depends on position. Then the second subsection \ref{Subsec: Section 2} will look at the OPE between two such operators. In such a scenario, the exchanged operator is annihilated by the same supercharges as the external operators and will therefore also be topological. Just as in a usual CFT, the quadratic Casimir can be used to construct blocs in this twisted conformal algebra with the difference that the exchanged operators all have vanishing eigenvalue. The quadratic Casimir acting on the external legs will therefore annihilate the correlator and give a constraint much like the Ward identities. The final subsection \ref{Subsec: section 3} will look at the explicit form of these equations and form linear differential constraints which match the superconformal Ward identities for 4 points and extend it to multipoint superconformal Ward identities.

% The following formalism is especially useful when considering multipoint correlators since it does not rely on finding all the superconformal invariants of the correlator. In particular, this explains why multipoint correlators have an increasing number of constraints.
\subsection{Topological sector}
Let us consider 1-dimensional $\mc{N}=2$ superconformal theory with supercharges 
\begin{align}
    &Q_a&&\bar{Q}^a&&S_a&&\bar{S}^a&&.
\end{align}
satisfying 
\begin{align}
    (Q_a)^\dag &= \bar{S}^a&(\bar{Q}^a)^\dag &= S_a,
\end{align}
and the 1/2-BPS supermultiplet $\Phi_{\Delta,\lambda}$ annihilated by $Q_1$ and $\bar{Q}^2$. The labels $\{\Delta,\lambda\}$ are the conformal and $su(2)$ weights respectively. The superprimary of this multiplet then satisfies 
\begin{align}
    &Q_1 \phi=\bar{Q}\phi = S_a \phi = \bar{S}^a \phi =0.
\end{align}
We can form the following fermionic operators
\begin{align}
    \mathbb{Q} &= Q_1+S_2 &\mathbb{Q}^\dag&=  \bar{Q}^2+\bar{S}^1
\end{align}
such that the following twisted 1d conformal algebra is $\mathbb{Q}/\mathbb{Q}^\dag$-exact:
\begin{align}
    \tilde{P}&= P+R^1_2 = \{\mathbb{Q},\bar{Q}^1\}\\
    \tilde{K}&= K-R^2_1 =(\tilde{P}^\dag)= -\{\mathbb{Q}^\dag,S_1\}\\
    \tilde{D}&= D -\frac 12 (R^1_1-R^2_2)= \frac 12\{\mathbb{Q},\mathbb{Q}^\dag\}
\end{align}
Where the details of the $su(2)$ and $sl(2)$ algebras are reminded in appendix \ref{Appendix: Simple lie algebras}. The particular choice of $\tilde{K}=(\tilde{P})^\dag$ seems natural to preserve some of the properties of the conformal group. $\Phi_{\Delta,\lambda}$ is therefore a topological operator with vanishing eigenvalue under $\tilde{D}$ 
\begin{align}
    \tilde{\Delta}=\Delta-\frac 12 \lambda =0.\label{Eq: vanishing of the eigenvalue}
\end{align}
The topological operators satisfy the following translation
\begin{align}
    \mathcal{O}_{top}(\tau) = \text{e}^{-\tau \tilde{P}} \mathcal{O}_{top}(0)  \text{e}^{\tau \tilde{P}}
\end{align}
So that when parametrising the translations in space and in the R-symmetry variable as :
\begin{align}
    \mathcal{O}_{\Delta,\lambda}(t,y) = \text{e}^{-P t - E_+ y}\mathcal{O}_{\Delta,\lambda}(0,0) \text{e}^{P t + E_+ y}
\end{align}
The topological limit corresponds to 
\begin{align}
    y_i\rightarrow t_i\label{Eq: topological limit}
\end{align}

\subsection{Topological OPE and quadratic Casimir}\label{Subsec: Section 2}
The operators above can be made topological in the limit \ref{Eq: topological limit}, we therefore have the constraint that the multipoint correlator of the 1/2-BPS operators considered above is a constant in this topological limit
\begin{align}
    \langle \mc{O}_{\Delta,2\Delta}(t_1,y_1)...\mc{O}_{\Delta,2\Delta}(t_n,y_n)\rangle \underset{y_i\rightarrow t_i}{=} \mathbb{F}_{0,n}
\end{align}
However, considering this topological twist of $sl(2)$ and $su(2)$ contains additional constraints from considering the OPE expansion in this topological sector. The topological operators have a natural OPE expansion in terms of other topological operators:
\begin{align}
    \hat{O}_1 \times \hat{O}_2&\sim \sum_i c_{12i}  O_i=0\\
    \mathbb{Q}(\hat{O}_1 \times \hat{O}_2) & \sim \sum_i c_{12i}\mathbb{Q} O_i=0\\
    O_i&=\hat{O}_i
\end{align}
Where we have used the fact that $\mathbb{Q}$ commutes with $P=\{\mathbb{Q},\bar{Q}_1\}$.
Just as in the usual conformal case,the twisted conformal algebra is naturally expressed in terms of conformal blocks which are eigenvalues under the quadratic Casimir\cite{Dolan:2000ut}. Let us consider the insertion of a resolution of the identity in a conformal correlator of operators with a topological limit
\begin{align}
    \langle O_1....O_n \rangle &=\langle O_1..O_j \sum_i\ket{O_i}\bra{O_i}..O_n \rangle
\end{align}
We consider the conformal Casimir of the twisted conformal algebra
\begin{align}
    \tilde{C}^{(2)}_{a_1...a_j} = \frac{1}{2}\{\sum_{i}\tilde{D}_{a_i},\sum_{i}\tilde{D}_{a_i}\}-\frac{1}{2}\{\sum_{i}\tilde{P}_{a_i},\sum_{i}\tilde{K}_{a_i}\}
\end{align}
The Casimir acting on the points $\{1,....,j\}$ of the correlator above gives:
\begin{align}
(\tilde{C}_{1...j}^{(2)}\langle O_1..O_j \sum_i\ket{O_i}\bra{O_i}..O_n \rangle)_{y_j\rightarrow t_j}&=\sum_i \langle O_1..O_j \hat{C}_{\Delta_i}O_i\rangle \langle O_i..O_n\rangle_{y_j\rightarrow t_j} \\
&=\sum_{i} \langle \hat{O}_1..\hat{O}_j \hat{C}_{\Delta_i}\hat{O}_i\rangle \langle \hat{O}_i..\hat{O}_n\rangle \\
&=\sum_i \tilde{\Delta}_i(\tilde{\Delta}_i-1) \langle \hat{O}_1..\hat{O}_j \hat{O}_i\rangle \langle \hat{O}_i..\hat{O}_n\rangle\\
&=0
\end{align}
Where we have used the fact that in the topological limit, only topological operators are exchanged, and they satisfy \ref{Eq: vanishing of the eigenvalue}. The quadratic Casimir equation therefore not only constrains the blocs, but the full correlator. Therefore, we know that the twisted quadratic Casimir acting on any number of points of a multipoint function will annihilate the correlator.\footnote{The attentice reader will worry about the order of limits, this is addressed in appendix \ref{Appendix: Linearity} and is the reason why the linearity of the differential operator is important.}
This set of quadratic differential equations for the correlator is naturally expressed in terms of spacetime and R-symmetry variables and is reminiscent of superconformal Ward identities.  These set of equations will lead to the superconformal Ward identities for the $n$-point correlator presented in the next subsection. \par 

\subsection{Multipoint Ward identities}\label{Subsec: section 3}
Let us consider the four-point correlator of 1/2-BPS operators of conformal dimension $\Delta$ considered above
\begin{align}
    \mc{A}_4 = \langle \mathcal{O}_{\Delta,2\Delta}(t_1,y_1) \mathcal{O}_{\Delta,2\Delta}(t_2,y_2) \mathcal{O}_{\Delta,2\Delta}(t_3,y_3) \mathcal{O}_{\Delta,2\Delta}(t_4,y_4)\rangle &=(\frac{y_{13}y_{24}}{t_{13}t_{24}})^{2\Delta}f_4(\chi,\rho)
\end{align}
where 
\begin{align}
    \chi &= \frac{t_{12}t_{34}}{t_{13}t_{24}}&\rho &= \frac{y_{12}y_{34}}{y_{13}y_{24}}.
\end{align}
The multipoint quadratic Casimir acts on the correlator as 
\begin{align}
    (C^{(2)}_{i}\mc{A}_4)|_{y_i=t_i} &= \tilde{\Delta}(\tilde{\Delta}-1)f_4(\chi_i,\chi_i)\\
    (C^{(2)}_{1,2}\mc{A}_4)|_{y_i=t_i}&=\chi^2 \left(\partial_\rho+(\chi-1) \partial_\rho^2+\partial_\chi+(\chi-1) \left(2 \partial_\rho\partial_\chi+\partial_\chi^2\right)\right)f_4(\chi,\rho)|_{\rho=\chi}\\
    (C^{(2)}_{2,3}\mc{A}_4)|_{y_i=t_i}&=-(\chi-1)^2 \left(\partial_\rho+\partial_\chi+\chi \left(\partial_\rho^2+2 \partial_\rho\partial_\chi+\partial_\chi^2\right)\right)f_4(\chi,\rho)|_{\rho=\chi}
\end{align}
The differential equations above can be recast in a linear form
\begin{align}
\left(\frac{C^{(2)}_{1,2}}{\chi}-\frac{C^{(2)}_{2,3}}{1-\chi}\right)f(\chi_i,\rho_i))|_{\rho=\chi}=\left( \partial_\chi+\partial_\rho\right)f(\chi_i,\rho_i))|_{\rho=\chi}
\end{align}
which gives the Ward identity for the four-point function:
\begin{align}
    \left( \partial_\chi+\partial_\rho\right)f(\chi_i,\rho_i))|_{\rho=\chi}=0
\end{align}
For higher point functions, we consider the correlator
\begin{align}
    \mc{A}_n =\langle \mathcal{O}_{\Delta,2\Delta}(t_1,y_1)...\mathcal{O}_{\Delta,2\Delta}(t_n,y_n)\rangle  = K_{n,\Delta,\lambda}(t_i,y_i)f_n(\chi_{i,n},\rho_{i,n})
\end{align}
where 
\begin{align}\label{prefactor}
    K_{n,\Delta,\lambda}(t_i,y_i)&=\left(\left(\frac{t_{1,n} t_{n-1,n}}{t_{1,n-1}}\right)^{n-2} \prod _{k=1}^{n-1} \frac{1}{t_{k,n}^2}\right)^{\Delta} \left(\left(\frac{y_{1,n} y_{n-1,n}}{y_{1,n-1}}\right)^{2-n} \prod _{k=1}^{n-1} y_{k,n}^2\right)^{\frac \lambda 2}
\end{align}
and
\begin{align}
    \chi_{i,n} &=\frac{t_{1,i+1}t_{n-1,n}}{t_{i+1,n}t_{1,n-1}} & \rho_{i,n} &=\frac{y_{1,i+1}y_{n-1,n}}{y_{i+1,n}y_{1,n-1}}.
\end{align}
There are now non-trivial constraints from $\frac{n(n-3)}{2}$ quadratic Casimirs which can be recast into $n-3$ linear differential equations. \par
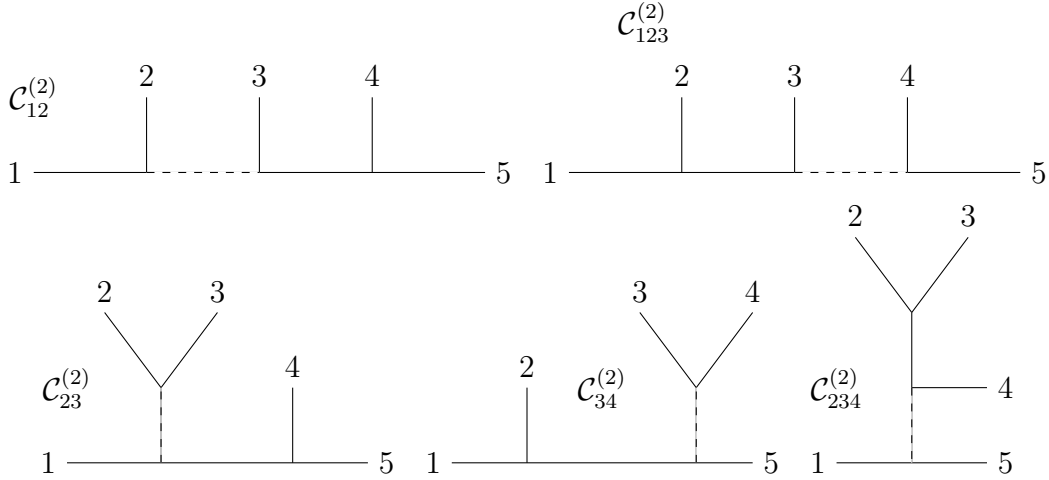
\begin{figure}
    \centering
    \begin{tikzpicture}
    \node[] at (0,1) {$\mc{C}_{12}^{(2)}$};
        \draw (0,0) -- (6,0);
        \draw (1.5,0) -- (1.5,1);
        \draw (3,0) -- (3,1);
        \draw (4.5,0) --(4.5,1);
        \draw[color=white] (1.5,0) -- (3,0);
        \draw[dashed] (1.5,0) -- (3,0);
        \node[anchor=east] at (0,0){$1$} ;
        \node[anchor=south] at (1.5,1) {$2$};
        \node[anchor=south] at (3,1) {$3$};
        \node[anchor=south] at (4.5,1)  {$4$};
        \node[anchor=west] at (6,0) {$5$};
    \end{tikzpicture}
        \begin{tikzpicture}
    \node[] at (1,2) {$\mc{C}_{123}^{(2)}$};
        \draw (0,0) -- (6,0);
        \draw (1.5,0) -- (1.5,1);
        \draw (3,0) -- (3,1);
        \draw (4.5,0) --(4.5,1);
        \draw[color=white] (3,0) -- (4.5,0);
        \draw[dashed] (3,0) -- (4.5,0);
        \node[anchor=east] at (0,0){$1$} ;
        \node[anchor=south] at (1.5,1) {$2$};
        \node[anchor=south] at (3,1) {$3$};
        \node[anchor=south] at (4.5,1)  {$4$};
        \node[anchor=west] at (6,0) {$5$};
    \end{tikzpicture}
    \begin{tikzpicture}
    \node[] at (1,1) {$\mc{C}_{23}^{(2)}$};
        \draw (1,0) -- (5,0);
        \draw (4,0) --(4,1);
        \draw (2.25,0) -- (2.25,1);
        \draw (2.25,1) -- (1.5,2);
        \draw (2.25,1) -- (3,2);
        \draw[color=white] (2.25,0) -- (2.25,1);
        \draw[dashed] (2.25,0) -- (2.25,1);
        \node[anchor=east] at (1,0){$1$} ;
        \node[anchor=south] at (1.5,2) {$2$};
        \node[anchor=south] at (3,2) {$3$};
        \node[anchor=south] at (4,1)  {$4$};
        \node[anchor=west] at (5,0) {$5$};
    \end{tikzpicture}
    \begin{tikzpicture}
    \node[] at (3,1) {$\mc{C}_{34}^{(2)}$};
        \draw (1,0) -- (5,0);
        \draw (2,0) --(2,1);
        \draw (4.25,0) -- (4.25,1);
        \draw (4.25,1) -- (3.5,2);
        \draw (4.25,1) -- (5,2);
        \draw[color=white] (4.25,0) -- (4.25,1);
        \draw[dashed] (4.25,0) -- (4.25,1);
        \node[anchor=east] at (1,0){$1$} ;
        \node[anchor=south] at (3.5,2) {$3$};
        \node[anchor=south] at (5,2) {$4$};
        \node[anchor=south] at (2,1)  {$2$};
        \node[anchor=west] at (5,0) {$5$};
    \end{tikzpicture}
    \begin{tikzpicture}
        \node[] at (3,0) {$\mc{C}_{234}^{(2)}$};
        \draw (4,0) -- (5,0);
        \draw (3,-1) -- (5,-1);
        \draw (4,-1) -- (4,1);
        \draw (4,1) -- (3.25,2);
        \draw (4,1) -- (4.75,2);
        \node[anchor=east] at (3,-1){$1$} ;
        \draw[color=white] (4,0) -- (4,-1);
        \draw[dashed] (4,0) -- (4,-1);
        \node[anchor=south] at (3.25,2) {$2$};
        \node[anchor=south] at (4.75,2) {$3$};
        \node[anchor=west] at (5,-1)  {$5$};
        \node[anchor=west] at (5,0) {$4$};
    \end{tikzpicture}
    \caption{OPE diagrams of the different configurations on which the twisted quadratic casimir acts on in the case of the 5-point function. The light grey lines represent the exchange of a topological operator with eigenvalue 0 under the twisted quadratic casimir leading to the vanishing of the correlator. For identical external operators, some of these are related by crossing, but the resulting equations are not dependent on that fact. }
    \label{Figure: OPE diagrams}
\end{figure}

 
For $n=5$, for example, we have:
\begin{align}
    \sum b^j_{a_i}\mathcal{C}^{(2)}_{a_i}  = \left(\partial_{\chi_j}+\partial_{\rho_j}\right)f_5 
\end{align}
where 
\begin{align}
    b^{1}_{1,2}&= \frac{1}{\chi_1}&b^{1}_{2,3}&=\frac{\chi_2-1}{(\chi_1-1) (\chi_1-\chi_2)}&b^1_{3,4}&=\frac{1}{1-\chi_1}&b^1_{1,2,3}&=0&b^1_{2,3,4}&=\frac{1}{\chi_1-1}\\
    b^{2}_{1,2}&=-\frac{1}{\chi_2}&b^{2}_{2,3}&=\frac{\chi_1}{\chi_2 (\chi_2-\chi_1)}&b^2_{3,4}&=\frac{1}{\chi_2-1}&b^2_{1,2,3}&=\frac{1}{\chi_2}&b^2_{2,3,4}&=0.
\end{align}
The action of a twisted quadratic Casimir on the correlator can be seen as the vanishing of the eigenvalue of the exchange operator in one of the OPE channels depicted in figure \ref{Figure: OPE diagrams}. 
This gives us the Ward identities:
\begin{align}
    \left(\partial_{\chi_1}+\partial_{\rho_1}\right) f_5 (\chi_i,\rho_i)|_{\rho_i=\chi_i}&=0&\left(\partial_{\chi_2}+\partial_{\rho_2}\right) f_5 (\chi_i,\rho_i)|_{\rho_i=\chi_i}&=0
\end{align}
The cases for higher points correlators are shown in the appendix \ref{Sec: Higher points}.\par 

This gives the general multipoint Ward identities for the operators with a topological sector with an $su(2)$ R-symmetry:
\begin{align}\label{Ward identity multipoint}
   \left(\partial_{\chi_j}+\partial_{\rho_j}\right) f_n (\chi_i,\rho_i)|_{\rho_i=\chi_i}=0 \qquad \forall j \in \{1,....,n-3\}
\end{align}

\section{An application: multipoint correlators on the 1/2 BPS Wilson line in $\mc{N}=4$ SYM}\label{Sec: Section 2}
In the case of the 1/2 BPS operators on the 1/2-BPS Wilson line defect in $\mc{N}=4$ SYM, the R-symmetry is $sp(4)\sim so(5)$ and we can define two pairs of nilpotent supercharges.\footnote{See, for example, the supercharges in equation (2.8) of \cite{Chester:2014fya,Chester:2014mea} and the footnote 13. This explains the similarity between the $su(2)$ case in the previous section and this one for $sp(4)$.} Using the notation in \cite{Ferrero:2023gnu,Ferrero:2023znz}, the 1/2-BPS operators are annihilated by the supercharges $\tilde{Q}_\alpha^a$ and the superconformal descendents are generated by the action of $Q_{\alpha a}$. As a consequence, there are two topological sectors. The four-point correlator can be conveniently written using the embedding formalism \cite{Dolan:2004mu}:
\begin{align}
    \langle \Phi_{\Delta,[k,0]}(t_1,Y_1)\Phi_{\Delta,[k,0]}(t_2,Y_2)\Phi_{\Delta,[k,0]}(t_3,Y_3)\Phi_{\Delta,[k,0]}(t_4,Y_4)\rangle = \frac{(Y_{12}Y_{34})^{k}}{(t_{12}t_{34})^{2\Delta}}f(\chi,\zeta\bar{\zeta},(1-\zeta)(1-\bar{\zeta}))
\end{align}
where the space time cross ratio is the same
\begin{align}
    \chi = \frac{t_{12}t_{34}}{t_{13}t_{24}}
\end{align}
and the R-symmetry cross ratios are
\begin{align}
    U&=\zeta\bar{\zeta}=\frac{Y_{12}Y_{34}}{Y_{13}Y_{24}}\\
    V&=(1-\zeta)(1-\bar{\zeta})=\frac{Y_{14}Y_{23}}{Y_{13}Y_{24}}
\end{align}
Which can be achieved by setting ($Y_{ij}=|\vec{y}_i-\vec{y}_j|^2$)
\begin{align}
    y_1&=y_4^{-1}=0&& y_3=(1,0,...)&&y_{2}=(\frac{\zeta+\bar{\zeta}}{2},\frac{1}{2} i (\zeta-\bar{\zeta}),0...).
\end{align}
In this notation, the shortening condition satisfied for both topological charges is\footnote{Note that between the $so(5)$ and $sp(4)$ notation, the two dynkin labels are swapped\cite{Ferrero:2023gnu}.} 
\begin{align}
    &\tilde{D}_i \Phi_{\Delta,[a,b]}=0&&\Delta=a&&b=0
\end{align}
which is satisfied by the 1/2-BPS multiplets related to the displacement multiplet (i.e those of the form $(\mathcal{D}_1)^k$). The two topological limits are
\begin{align}
    \zeta \rightarrow \chi&&\bar{\zeta} \rightarrow \chi
\end{align}
and there is a symmetry between the barred and unbarred variables. The Ward identities were derived for this case in \cite{Liendo:2016ymz} and read
\begin{align}
    (\frac 12 \partial_\chi+\partial_\zeta)f|_{\zeta\rightarrow\chi}=0\label{Ward 4 pt}.
\end{align}
or equivalently as\cite{Barrat:2021tpn}
\begin{align}
    (\frac 12 \partial_\chi +\alpha \partial_r-(1-\alpha)\partial_s)f(\chi,r,s)|_{r\rightarrow \alpha \chi,s\rightarrow (1-\alpha)(1-\chi)}=0
\end{align}
where the function is explicitly symmetric in $\zeta,\bar{\zeta}$. In this formalism, it is clear that the $sp(4)$ algebra can be seen as two $su(2)$ algebras, one rotating and the other projecting \cite{Liendo:2015cgi} while the symmetry between $\zeta$ and $\bar{\zeta}$ is the automorphism between the two copies of $su(2)$.\par  For the higher-point functions, the correlator will in general depend on $n-3$ space-time cross ratios and $2(n-3)+\frac{(n-4)(n-3)}{2}$ R-symmetry cross-ratios. These map conveniently to the $\chi_i$ and $\zeta_i, \bar{\zeta}_i, \eta_{ij}$ variables where we keep the automorphism by imposing the constraint $\bar{\eta}_{ij}(\zeta_i-\zeta_j)=\eta_{ij}(\bar{\zeta}_i-\bar{\zeta}_j)$ for the dummy variable $\bar{\eta}_{ij}$. For higher points, some of these R-symmetry cross ratios become degenerate which reduces them to $3(n-3)-1$ variables. In terms of the Mandelstam variables $r_i,s_i,t_{ij}$, this amounts to keeping the definitions for $r$ and $s$ and having $n-4$ independent $t_{ij}$ variables described by $\frac{(n-4)(n-3)}{2}$ partially redundant variables. This works since each Ward identity depends on $n-4$ of these variables.\footnote{One way to see this is in analogy to the Mellin variables $\gamma_{ij}$ as explained in \cite{Mack:2009mi}, the redundant description allows for a simpler formulation. If needed, one could fix one variable in term of the others.}
Explicitly, we consider the $n-$point function
\begin{align}
    \langle\Phi_{k,[k,0]}(t_1,Y_1)... \Phi_{k,[k,0]}(t_n,Y_n)\rangle = K_{n,k,k}(t_i,Y_i)f(\chi_i,r_i,s_i,t_{ij})
\end{align}
where the prefactor is defined in \ref{prefactor} and the cross ratios are given by
\begin{align}
    \chi_{i,n} &=\frac{t_{1,i+1}t_{n-1,n}}{t_{i+1,n}t_{1,n-1}}& t_{i,j,n}&= \eta_{ij}(\bar{\zeta}_i-\bar{\zeta}_j)= \bar{\eta}_{ij}(\zeta_i-\zeta_j)= \frac{Y_{1,n} Y_{n-1,n} Y_{i+1,j+1}}{Y_{1,n-1} Y_{i+1,n} Y_{j+1,6}} \\ r_{i,n} &= \zeta_i\bar{\zeta}_i=\frac{Y_{1,i+1}Y_{n-1,n}}{Y_{i+1,n}Y_{1,n-1}}&s_{i,n} &= (1-\zeta_i)(1-\bar{\zeta}_i)=\frac{Y_{1,n}Y_{i+1,n-1}}{Y_{i+1,n}Y_{1,n-1}}  
\end{align}
The generalisation of the Ward identity \ref{Ward 4 pt} of the same type as \ref{Ward identity multipoint} would then be
\begin{align}
%   &(\frac 12 \partial_{\chi_i} +\partial_{\zeta_i}+\sum_j\partial_{\eta_{ij}})f(\chi_i,\zeta_i \bar{\zeta_i},(1-\zeta_i)(1-\bar{\zeta_i}),\eta_{ij}(\bar{\zeta}_i-\bar{\zeta}_j)))|_{\zeta_i\rightarrow \chi_i,\eta_{ij}\rightarrow\chi_i-\chi_j}=0\\
    &(\frac 12 \partial_{\chi_{i^*}} +\partial_{\zeta_{i^*}})f(\chi_i,\zeta_i \bar{\zeta_i},(1-\zeta_i)(1-\bar{\zeta_i}),\bar{\eta}_{ij}(\zeta_i-\zeta_j)))|_{\zeta_i\rightarrow \chi_i,\eta_{ij}\rightarrow\chi_i-\chi_j}=0
\end{align}
where the constraint $\bar{\eta}_{ij}(\zeta_i-\zeta_j)=\eta_{ij}(\bar{\zeta}_i-\bar{\zeta}_j)$ restores the symmetry between the barred and unbarred variables.
%\begin{align}
%   &(\frac 12 \partial_{\chi_i} +\partial_{\zeta_i}+\sum_j\partial_{\eta_{ij}})f(\chi_i,\zeta_i \bar{\zeta_i},(1-\zeta_i)(1-\bar{\zeta_i}),\eta_{ij}(\bar{\zeta}_i-\bar{\zeta}_j)))|_{\zeta_i\rightarrow \chi_i,\eta_{ij}\rightarrow\chi_i-\chi_j}=0\\
%    &(\frac 12 \partial_{\chi_i} +\partial_{\bar{\zeta}_i})f(\chi_i,\zeta_i \bar{\zeta_i},(1-\zeta_i)(1-\bar{\zeta_i}),\bar{\eta_{ij}}(\zeta_i-\zeta_j)))|_{\bar{\zeta}_i\rightarrow \chi_i,\bar{\eta}_{ij}\rightarrow\chi_i-\chi_j}=0
%\end{align}
%\begin{align}
 %  (\frac 12 \partial_{\chi_i} +\partial_{\bar{\zeta}_i}+\sum_j\partial_{\bar{\eta}_{ij}})f(\chi_i,\zeta_i \bar{\zeta_i},(1-\zeta_i)(1-\bar{\zeta_i}),\bar{\eta}_{ij}(\zeta_i-\zeta_j)))|_{\bar{\zeta}_i\rightarrow \chi_i,\bar{\eta}_{ij}\rightarrow\chi_i-\chi_j}=0
%\end{align}
However, when recasting the Casimir in its linear form, this can only be achieved by setting $\bar{\zeta}_{i^*}\rightarrow 1$.
Rewriting this in terms of the $r_i$, $s_i$ and $t_{ij}$ variables used in \cite{Barrat:2021tpn} is then easy which gives $n-3$ linear superconformal Ward identities
\begin{align}
    \lim_{\alpha_{i^*}\rightarrow 1}(\frac 12\partial_{\chi_{i^*}} +\alpha_{i^*} \partial_{r_{i^*}}-(1-\alpha_{i^*} )\partial_{s_{i^*}}+\sum_j (\alpha_{i^*}-\alpha_j )\partial_{t_{{i^*}j}})f(\chi_i,r_i,s_i,t_{ij}))\raisebox{-.2cm}{$\Biggr|$}_{\raisebox{+.2cm}{$\begin{subarray}{l}
				r_i\rightarrow \alpha_i \chi_i \\s_i\rightarrow (1-\alpha_i)(1-\chi_i)\\t_{ij}\rightarrow (\alpha_i-\alpha_j)(\chi_i-\chi_j)
			\end{subarray}$}}=0.
\end{align}
where the sum of these terms then gives a weaker form of the superconformal Ward identity for multipoints conjectured in \cite{Barrat:2021tpn}:
\begin{align}
    	\lim_{\alpha_i\rightarrow 1}\sum_{k=1}^{n-3} \left(\frac{1}{2}\partial_{\chi_k}+\alpha_k\partial_{r_k}-(1-\alpha_k)\partial_{s_k} \right)\mathcal{A}_{\Delta_1...\Delta_n}\raisebox{-.2cm}{$\Biggr|$}_{\raisebox{+.2cm}{$\begin{subarray}{l}
				r_i\rightarrow \alpha_i \chi_i \\s_i\rightarrow (1-\alpha_i)(1-\chi_i)\\t_{ij}\rightarrow (\alpha_i-\alpha_j)(\chi_i-\chi_j)
			\end{subarray}$}} =  0
\end{align}
where we have used the fact that 
\begin{align}
    \sum_{ij}(\alpha_i-\alpha_j)\partial_{t_{ij}}=0.\label{Eq sum tij}
\end{align}
It would be interesting to see if the linearising of the quadratic Casimir can directly give the sum conjectured in \cite{Barrat:2021tpn}.
%For example, on the 6-point function of the displacement, the 3 Ward identities are 
%\begin{align}
%    &(\frac 12\partial_{\chi_i} +\alpha_i \partial_{r_i}-(1-\alpha_i )\partial_{s_i}+\sum_j (\alpha_i-\alpha_j )\partial_{t_{ij}})f(\chi_i,r_i,s_i,t_{ij}))|_{Topo}=0
    %\\
    %&(\frac 12\partial_{\chi_2} +\alpha_2 \partial_{r_2}-(1-\alpha_2 )\partial_{s_2}+\sum_j (\alpha_2-\alpha_j )\partial_{t_{2j}})f(\chi_i,r_i,s_i,t_{ij}))|_{Topo}=0\\
   %&(\frac 12\partial_{\chi_3} +\alpha_3 \partial_{r_3}-(1-\alpha_3 )\partial_{s_3}+\sum_j (\alpha_3-\alpha_j )\partial_{t_{3j}})f(\chi_i,r_i,s_i,t_{ij}))|_{Topo}=0
%\end{align}
%where $Topo$ indicates the topological limit
%\begin{align}
%    r_i&\rightarrow\alpha_i \chi_i&s_i&\rightarrow (1-\alpha_i)(1-\chi_i)&t_{ij}&\rightarrow(\alpha_i-\alpha_j)(\chi_i-\chi_j)
%\end{align}
%and $t_{ij}$ are symmetric. 
%Notice that the condition \ref{Eq sum tij}  is satisfied
%\begin{align}((\alpha_1-\alpha_2)+(\alpha_2-\alpha_1))\partial_{t_{12}}+((\alpha_1-\alpha_3)+(\alpha_3-\alpha_1))\partial_{t_{13}}+((\alpha_2-\alpha_3)+(\alpha_3-\alpha_2))\partial_{t_{23}}=0\nonumber
%\end{align}
These annihilate the 5- and 6-point superblocks derived in \cite{Barrat:2024nod, Bliard:2023zpe,Peveri:2023qip} as well as the strong coupling results in \cite{Bliard:2023zpe}. 
These identities can only be summed by having a coincident limit so give a weaker version of the Ward Identity in \cite{Barrat:2021tpn}. However, these still contain $n-3$ constraints so can probably be rewritten in a more convenient form. The increase on the number of constraints is clear from the structure as well as the derivation of the Ward identity. There are $n-3$ Ward identities corresponding to the different linearisations of the quadratic Casimir equations. For each of these, there are $n-4$ variables ($\bar{\zeta}$ in the first formulation and $\alpha$ in the second) that are unfixed from the topological limit. This being said, it is unclear how these  combine and if how many constraints these Ward identities give for a generic $n-$point correlator though it was conjectured that the number of independent constraints should grow as $C \frac{n}{2}$\cite{Barrat:2021tpn} at large $n$ where $C$ is the Catalan number.\footnote{The author thanks Julien Barrat for pointing this out.}
\section{Conclusion}
In this note, the $n$-point correlators of supersymmetric operators with a topological limit were studied. These are known to satisfy topological constraints, that is, the correlator is constant in the topological limit. To see this, one can construct a twisted conformal algebra which is exact under the supercharges annihilating the superprimary. Therefore, when acting on the correlator with the twisted translations, one is effectively acting with a supercharges giving 0. However, one can further constrain these correlators by considering the operator product expansion (OPE) between two such operators. In particular, the topological condition satisfied by the primary field $\Phi$ will also be satisfied by these exchanged operators. As a consequence, when acting with the generators of the twisted conformal algebra on the external legs of an OPE diagram, one annihilates the whole multipoint function. The corresponding twisted quadratic Casimir acting on one point only gives the shortening condition of $\Phi$. However, the remaining quadratic Casimirs acting on the correlator lead to non-trivial linear differential equations involving both the R-symmetry and the spacetime coordinates of the correlator. Recast in its linear form for the simplest case of $su(2)$ R-symmetry, this gives the Ward identities
\begin{align}(\partial_{\chi_i}+\partial_{\rho_i})f(\chi_i,\rho_i)|_{\rho_i \rightarrow \chi_i}=0.
\end{align}
One of the questions one could ask in the usual derivation of multipoint Ward identities such as in \cite{Barrat:2021tpn} was why there should be an increasing number of constraints for an increasing number of external points while the number of supersymmetries remained fixed. In particular, when deriving them, one would first need to fix the superconformal invariants before analysing the spurious poles from the variations in supersymmetry. \par Rather, in this note, the source of these constraints are clear; when acting on an $n-$point function with the twisted conformal Casimir, one isolates the eigenvalue of the corresponding exchanged topological operator which, by definition, must be zero. For an increasing number of external legs, there is an increasing number of internal exchanged operators which in turn leads to an increasing number of constraints from the Ward identities.\par 
This analysis finds its applications in the plethora of superconformal theories and superconformal defects, such as \cite{Liendo:2016ymz,Bianchi:2020hsz,Correa:2021sky,Chester:2014mea}. In particular, this provides a method to prove the multipoint Ward identities conjectured in \cite{Barrat:2021tpn} which are crucial to the construction of the superconformal blocks needed for the analysis of analytic or numerical bootstrap such as in \cite{Lemos:2015awa,Beem:2016wfs}. In this light, the note explores the equivalent multipoint Ward identities for correlators of insertions on the 1/2 BPS Wilson line in $\mc{N}=4$ SYM which are
\begin{align}
(\frac 12\partial_{\chi_{i^*}} +\partial_{\zeta_{i^*}})f(\chi_i,\zeta_i\bar{\zeta}_i,(1-\zeta_i)(1-\bar{\zeta}_i),(\zeta_i-\zeta_j)\bar{\eta_{ij}})\raisebox{-.2cm}{$\Biggr|$}_{\raisebox{+.2cm}{$\begin{subarray}{l}
				\zeta_i\rightarrow \chi_i \\\eta_{ij}\rightarrow (\chi_i-\chi_j)\\\bar{\zeta}_{i^*}\rightarrow 1
			\end{subarray}$}}=0.
\end{align}
These can be rewritten as
\begin{align}
    \lim_{\alpha_{i}\rightarrow 1}(\frac 12\partial_{\chi_i} +\alpha_i \partial_{r_i}-(1-\alpha_i )\partial_{s_i}+\sum_j (\alpha_i-\alpha_j )\partial_{t_{ij}})f(\chi_i,r_i,s_i,t_{ij}))\raisebox{-.2cm}{$\Biggr|$}_{\raisebox{+.2cm}{$\begin{subarray}{l}
				r_i\rightarrow \alpha_i \chi_i \\s_i\rightarrow (1-\alpha_i)(1-\chi_i)\\t_{ij}\rightarrow (\alpha_i-\alpha_j)(\chi_i-\chi_j)
			\end{subarray}$}}=0
\end{align}
and can be seen either as a rewriting or a limit of the multipoint Ward identities in \cite{Barrat:2021tpn}. This paves the way for a derivation using the superspace formalism and indicates that there would indeed be poles at the coincident point $\zeta_i=\chi_i$ While being technically more challenging, this would be extremely interesting since the resulting construction then gives crucial information for the descendents of the operators which is not given by the bosonic correlator above.\footnote{At the time of publication, work is being done in this direction\cite{multipoints:sophie}}.\par 
This analysis provides an efficient way to derive superconformal Ward identities, avoiding the complications of grassmann coordinates while focusing only on the space-time and R-symmetry coordinates. Furthemore, this formalism seems widely applicable to systems with a topological sector and a topologically twisted algebra. Beyond the other line defect cases listed in the classification of \cite{Agmon:2020pde}, there are several higher-dimensional examples that could use this same method. In particular, there is a full two-dimensional $Q-$exact conformal algebra in the case of $\mc{N}=4$ SYM so it should follow that the same method can be used to derive the SCWI for the 1/2-BPS surface defect in this theory. In $d=3$, the configuration of topological operators in ABJM \cite{Guerrini:2023rdw} lying in a circle around the Wilson line could be studied in this setting. Additionally, there is the symmetry between the 1-dimensional and codimension-1 defect in $\mc{N}=4$ SYM\cite{Liendo:2016ymz} which indicates that there should be a similar derivation for this $p=3$ defect.  The only uncertain aspect when extending this analysis to higher dimensions is whether the Casimir equations can be recast into a linear form. One example in which this is clearly not the case is the four point function with two 1/2-BPS topological operators and two general longs such as those studied in \cite{Ferrero:2023znz}. Beyond the absence of a topological limit, the condition from the twisted Casimir acting on points 1 and 2 is quadratic and cannot be linearised. As a consequence the order of limits discussed in the main text plays a role and there are no constraints akin to the SCWI. The application to higher-dimensions is a logical extension to this work and it will be interesting to see if the same simplifications occur.  \par 
In a system where this method can be applied, the resulting SCWI can be used to fix the form of the superconformal blocs in the OPE expansion of the $n-$point correlator of the superconformal primary of the topological multiplet, such as is \cite{Dolan:2004iy,Bianchi:2020hsz,Liendo:2018ukf,Bissi:2015qoa}. For 5 and 6-point functions of the displacement multiplet in the 1/2-BPS Wilson line defect in $\mc{N}=4$ SYM, this was done in \cite{Bliard:2023zpe} in anticipation using the conjecture of \cite{Barrat:2021tpn}. The form of these superconformal blocs as well as that of the correlator constrained by the SCWI is then enough to study the system using many other techniques; If the correlator is known through explicit computations, the unmixed CFT data can then be obtained. Likewise, the numerical bootstrap can be used with these blocs to obtain bounds on the CFT data \cite{Liendo:2018ukf,Antunes:2023dlk}. Perturbatively, the analytical bootstrap can be used to find the form of the correlator from this expansion, an ansatz and the crossing equation. The case-by-case use of these SCWI to fix the form of the superconformal blocs in different theories is a rich area which supports the recent interest in higher-point correlators.\par 
This analysis facilitates the study of multipoint correlators, particularly in the context of line defects. The analysis of these multipoint correlators can then shed light on several aspects including finite deformations of defects, resolving the mixing problem and determining the complete CFT data of the defect theory.
\section*{Acknowlegements}
The author is grateful to Aleix Gimenei-Grau for sharing unpublished notes which led to the initial inspiration, and to Miguel Paulos, Philine Van Vliet, Adam Chalabi and Maxime Trepanier for very useful discussions. The author acknowledges an ongoing project \cite{multipoints:sophie} whose scope is complementary to this work and thanks the authors for discussions. This work was funded by the European Union (ERC, FUNBOOTS, project number 101043588). Views and opinions expressed are however those of the author(s) only and do not necessarily reflect those of the European Union or the European Research Council. Neither the European Union nor the granting authority can be held responsible for them.\newpage
\appendix
\section{Simple Lie algebras}\label{Appendix: Simple lie algebras}
\subsection*{$SL(2,\mathbb{C})$}
The Conformal algebra of the 1d-defect is $SL(2,\mathbb{C})$, generated by $\{P,K,D\}$ satisfying the commutation relations
\begin{align}
    [D,P]&=P&[D,K]&=-K&[K,P]&=2D
\end{align}
We can write these as differential operators acting on a field 
\begin{align}
    \Phi_\Delta(t) &= \text{e}^{-P t}  \Phi_\Delta(0)\\
    P &= -\partial_t\\
    K&=-t^2\partial_t-2\Delta t \\
    D &= t\partial_t+\Delta
\end{align}
The quadratic Casimir is then written as :
\begin{align}
    \mathcal{C}^{(2)}_{a_1,...,a_n} = \frac{1}{2}\{\sum_i D_i,\sum_i D_i\}-\frac{1}{2}\{\sum_i P_i,\sum_i K_i\}
\end{align}
Explicitly, this gives the Casimir eigenvalue and 2-point differential operator
\begin{align}
    \mathcal{C}^{(1)}_{a_1}&= \Delta(\Delta-1)\\
    \mathcal{C}^{(2)}_{1,2}&= (\Delta_1+\Delta_2)(\Delta_1+\Delta_2-1)+(t_1-t_2)\left(2\Delta_2\partial_{t_1}-2\Delta_1\partial_{t_2}+(t_1-t_2)\partial_{t_1}\partial_{t_2}\right)
\end{align}
Which give the usual differential equation for the 4-point conformal bloc:
\begin{align}
    K(t_i)^{-1}\mathcal{C}^{(2)}_{1,2}K(t_i)G_\Delta (\chi)&= \Delta(\Delta-1)G_\Delta(\chi)\\
    G_{\Delta}(\chi)&= \chi^\Delta {}_2F_1(\Delta,\Delta,2\Delta,\chi)
\end{align}
where we factor use the prefactor 
\begin{align}
    K(t_i)=\frac{1}{(t_{12}t_{23})^{2\Delta}}
\end{align}
for two identical operators $\Delta_1=\Delta_2=\Delta$
\subsection*{$SU(2)$}
In the case of SU(2) in the Chevalley basis, we have the commutation relations:
\begin{align}
    [H_0,E_\pm]&= \pm E_\pm &[E_+,E_-]&= 2H_0
\end{align}
For an operator of spin J, we have the following differential action of the elements of the algebra
\begin{align}
\Phi_{\lambda}(y)&=\text{e}^{-y E_-}\Phi_{\lambda}(0)\\
    E_m &= -\partial_y\\
    E_+ &= y^2\partial_y-\lambda y \\
    H_0 &= y \partial_y +\frac{\lambda}{2}
\end{align}
The quadratic Casimir in this basis is\footnote{Note that the Cartan is scaled by 2 compared to the Chevalley basis, and that the Casimir differs from the sl(2) case because of the rescaling of the Dilation operator $D\rightarrow-iD$ which is standard in $1d$.} 
\begin{align}
    \mathcal{C}^{(2)}_{su(2),a_i} = \frac{1}{2}\{\sum_i H^i_0,\sum_iH^i_0\}+\frac 12 \{\sum_iE^i_-,\sum_iE^i_+\}
\end{align}
The eigenfunctions of the 2-point Casimir, the su(2) blocs for 4-point functions, are then found to be (once the prefactor \ref{prefactor} is factored out):
\begin{align}
    \mathcal{C}^{(2)}_{su(2),1,2} H_\lambda &= \frac \lambda 2(\frac \lambda 2+1)H_\lambda \\
    H_\lambda&=\rho^{J}{}_2F_1(-\frac \lambda 2,-\frac \lambda 2,- \lambda ,\rho)
\end{align}
\subsection*{Twisted conformal algebra}
Combining these two previous sections, we can write an operator of spin $J=\frac \lambda 2$ and weight $\Delta$ as:
\begin{align}
    \Phi_{\Delta,\lambda}(t,y)=\text{e}^{-t P-y E_-}\Phi_{\Delta,\lambda}(0,0)
\end{align}
Where $\Phi_{\Delta,\lambda}$ is a primary of both the $sl(2)$ and the $su(2)$ and therefore satisfies:
\begin{align}
    K \Phi_{\Delta,\lambda}(0,0)&=(D-\Delta) \Phi_{\Delta,\lambda}(0,0)=0\\
    E_+ \Phi_{\Delta,\lambda}(0,0)&= (H_0-\frac \lambda 2) \Phi_{\Delta,J}(0,0)=0
\end{align}
The twisted conformal algebra can then be written as :
\begin{align}
    \tilde{P} &= P+E_-\\
    \tilde{K} &= K-E_+ \\
    \tilde{D} &=D-H_0
\end{align}
Since the $su(2)$ and the $sl(2)$ commute, we can just add the operators to form the differential operators of the twisted algebra and the twisted quadratic Casimir
\begin{align}
    \tilde{\mc{C}}_{a_i}^{(2)}&=\frac{1}{2}\{\sum_i \tilde{D}_i,\sum_i \tilde{D}_i\}-\frac{1}{2}\{\sum_i \tilde{P}_i,\sum_i \tilde{K}_i\}
\end{align}
this gives the Casimir Eigenvalue 
\begin{align}
    \mathcal{C}^{(2)}_{a_1}&= \tilde{\Delta}(\tilde{\Delta}-1)\\
    \tilde{\Delta}&=\Delta-\frac \lambda 2
\end{align}
and the higher-point Casimir acting on multipoint correlators of topological operators give the corresponding Ward identities in the main text.
%\section{$SU(2)\in SU(N)$}
%\section{$su(4)$}
%\section{Differential operators and equations}
\section{Linearity of the Casimir operator}\label{Appendix: Linearity}
Near the topological limit, the bloc decomposition can be separated into 2 parts
\begin{align}
   \sum_\Delta G_\Delta(\chi,\rho) \underset{\rho\rightarrow \chi}{=} \sum_{\text{topo}} G_{\tilde{\Delta}}(\chi,\chi)+\sum_{\text{not topo}} (\rho-\chi)G_{\tilde{\Delta}}(\chi,\chi)+O((\rho-\chi)^2)
\end{align}
In the topological limit, the second term trivially disappears. When acting on the OPE with a linear differential operator made from the twisted conformal Casimirs
\begin{align}
    \mc{L} = \sum_{a_i}\mathcal{C}^{(2)}_{a_i}
\end{align}
we can write it as:
\begin{align}
   &\mc{L}\sum_\Delta G_\Delta(\chi,\rho) \underset{\rho\rightarrow \chi}{=} \\
   &\quad \sum_{\text{topo}} \mc{L}G_{\tilde{\Delta}}(\chi,\chi)+\sum_{\text{not topo}} \mc{L}(\rho-\chi)G_{\tilde{\Delta}}(\chi,\chi)+\sum_{\text{not topo}} (\rho-\chi)\mc{L}G_{\tilde{\Delta}}(\chi,\chi)+O((\rho-\chi)^2)\nonumber
\end{align}
The first term has 0-eigenvalue under the quadratic Casimir (\ref{Eq: vanishing of the eigenvalue}), the $(\rho-\chi)$ term is annihilated by the quadratic Casimir (by definition, the quadratic Casimir commutes with the twisted translations.) so that in the topological limit, all three terms vanish for a linear differential operator. Thankfully, such an operator can be constructed as a linear combination quadratic Casimirs. We therefore have the equation:
\begin{align}
    \sum_{a_i} b_{a_i} \tilde{C}^{(2)}_{a_i} \langle O_1....O_n \rangle \underset{\rho\rightarrow \chi}{=} 0 
\end{align}
%In terms of the OPE, this is understood in a similar way: OPE between two operators can be written as:
%\begin{align}
%    \Phi_{\Delta,2\Delta}(t_1,y_1) \times \Phi_{\Delta,2\Delta}(t_2,y_2)  \sim \sum_{top} O_i +(t_{1,2}-y_{1,2})\sum_{top} O
%\end{align}
%In the topological limit, the second term trivially disappears. When acting on the OPE with a linear differential operator made from the twisted conformal Casimirs
%\begin{align}
%    \mc{L} = \sum_{a_i}\mathcal{C}^{(2)}_{a_i}
%\end{align}
%we can write it as:
%\begin{align}
%    \mc{L}\left(\Phi_{\Delta,2\Delta}(t_1,y_1) \times \Phi_{\Delta,2\Delta}(t_2,y_2) \right) \sim \sum_{top} \mc{L}O_i +\mc{L}\left((t_{1,2}-y_{1,2})\sum_{non-top} O\right)
%\end{align}
%The twisted Casimirs commute with the topological distance so we have:
%\begin{align}
%    \lim_{y_i\rightarrow t_i}\mc{L}\left(\Phi_{\Delta,2\Delta}(t_1,y_1) \times \Phi_{\Delta,2\Delta}(t_2,y_2) \right) =0
%\end{align}
In particular, this means that the topological limit of the linearised quadratic Casimir acting on a multipoint correlator of operators with a topological limit vanishes. This gives the superconformal Ward identity constraints on the $n-$point correlators of the superprimary in \ref{Subsec: section 3}.

\section{Higher-points}\label{Sec: Higher points}
For higher points, the expressions quickly become overwhelming. It is convenient to define the distance:
\begin{align}
    \chi_i(j,k) = \frac{1}{\chi_i-\chi_j}- \frac{1}{\chi_i-\chi_k}
\end{align}
where we extend the definition of $\chi$ to 
\begin{align}
    \chi_0 &=0&\chi_{n-2}&=1&\chi_{n-1} &= \infty 
\end{align}
so that we have, for example:
\begin{align}
    \chi_i(0,n-1) = \frac{1}{\chi_i}
\end{align}
In this language we have a list of the coefficients leading to the linear combination of quadratic Casimirs resented in the main text
\subsection*{$n=6$ points}
\begin{align}
    &b^1_{1,2}=-\chi_1(5,0)&&b^1_{2,3}=\chi_1(2,3)&&b^1_{3,4}=-\chi_1(3,4)&&b^1_{4,5}=0&b^1_{1,2,3}= 0\nonumber\\
    &b^1_{2,3,4}=\chi_1(3,4)&&b^1_{3,4,5}=\chi_1(4,5)&&b^1_{1,2,3,4}=0&&b^1_{2,3,4,5}=-\chi_1(4,5)
\end{align}
\begin{align}
    &b^2_{1,2}=\chi_2(5,0)&&b^2_{2,3}=-\chi_2(0,1)&&b^2_{3,4}=\chi_2(3,4)&&b^2_{4,5}=-\chi_2(4,5)&b^1_{1,2,3}=0\nonumber\\
    &b^2_{2,3,4}=0&&b^2_{3,4,5}=\chi_1(4,5)&&b^1_{1,2,3,4}=0&&b^1_{2,3,4,5}=0
\end{align}
\begin{align}
    &b^3_{1,2}=0&&b^3_{2,3}=\chi_3(0,1)&&b^3_{3,4}=\chi_3(1,2)&&b^3_{4,5}=\chi_3(4,5)&b^3_{1,2,3}=-\chi_3(0,5)\nonumber\\
    &b^3_{2,3,4}=-\chi_3(0,1)&&b^3_{3,4,5}=0&&b^3_{1,2,3,4}=\chi_3(0,5)&&b^3_{2,3,4,5}=0
\end{align}
\subsection*{$n=7$ points}
\begin{align}
    &b^1_{1,2}=-\chi_1(6,0)&&b^1_{2,3}=\chi_1(2,3)&&b^1_{3,4}=-\chi_1(3,4)&&b^1_{4,5}=0&&b^1_{5,6}=0\nonumber\\
    &b^1_{1,2,3}= 0&&b^1_{2,3,4}=\chi_1(3,4)&&b^1_{3,4,5}=-\chi_1(4,5)&&b^1_{4,5,6}=0&&b^1_{1,2,3,4}=0&\nonumber\\
    &b^1_{2,3,4,5}=\chi_1(4,5)&&b^1_{3,4,5,6}=-\chi_1(5,6)&&b^1_{1,2,3,4,5}=0&&b^1_{2,3,4,5,6}=\chi_1(5,6)
\end{align}
\begin{align}
    &b^2_{1,2}=\chi_2(6,0)&&b^2_{2,3}=\chi_2(0,1)&&b^2_{3,4}=\chi_1(3,4)&&b^2_{4,5}=\chi_2(4,5)&&b^1_{5,6}=0\nonumber\\
    &b^2_{1,2,3}= -\chi_2(6,0)&&b^2_{2,3,4}=0&&b^2_{3,4,5}=-\chi_2(4,5)&&b^2_{4,5,6}=-\chi_2(5,6)&&b^2_{1,2,3,4}=0&\nonumber\\
    &b^2_{2,3,4,5}=0&&b^2_{3,4,5,6}=\chi_2(5,6)&&b^2_{1,2,3,4,5}=0&&b^2_{2,3,4,5,6}=0
\end{align}
\begin{align}
    &b^3_{1,2}=0&&b^3_{2,3}=\chi_3(0,1)&&b^3_{3,4}=-\chi_3(1,2)&&b^3_{4,5}=\chi_3(4,5)&&b^3_{5,6}=\chi_3(5,6)\nonumber\\
    &b^3_{1,2,3}= \chi_3(6,0)&&b^3_{2,3,4}=\chi_3(0,1)&&b^3_{3,4,5}=0&&b^3_{4,5,6}=-\chi_3(5,6)&&b^3_{1,2,3,4}=-\chi_3(6,0)&\nonumber\\
    &b^3_{2,3,4,5}=0&&b^3_{3,4,5,6}=0&&b^3_{1,2,3,4,5}=0&&b^3_{2,3,4,5,6}=0
\end{align}
\begin{align}
    &b^4_{1,2}=0&&b^4_{2,3}=0&&b^4_{3,4}=\chi_4(1,2)&&b^4_{4,5}=-\chi_4(2,3)&&b^4_{5,6}=\chi_4(5,6)\nonumber\\
    &b^4_{1,2,3}= 0&&b^4_{2,3,4}=\chi_4(0,1)&&b^4_{3,4,5}=-\chi_4(1,2)&&b^4_{4,5,6}=0&&b^4_{1,2,3,4}=\chi_4(6,0)&\nonumber\\
    &b^4_{2,3,4,5}=\chi_4(0,1)&&b^4_{3,4,5,6}=0&&b^4_{1,2,3,4,5}=-\chi_4(6,0)&&b^4_{2,3,4,5,6}=0
\end{align}
\subsection*{$n\geq 8$ points}
Until $n=12$, the same Ward identities are found. The relevant coefficients still have a simple expression in terms of $\chi_i(j,k)$ but are not particularly enlightening so have not been included in this note.
%\section{sp(4) topological charges}
%In the notation of \cite{Ferrero:2023gnu}, we can form the topological charges:
%\begin{align}
 %   \mathbb{Q} = \tilde{Q}_{\alpha}^1+\xi \tilde{S}_2^\beta &&\mathbb{Q}^\dag = \tilde{Q}_{\alpha}^2+\xi^{-1}\tilde{S}_1^\beta 
%\end{align}
%where the choice of $\alpha\neq \beta$ is arbitrary, we set $\{\alpha,\beta\}=\{1,2\}$ and $\xi$ is an arbitrary phase. The twisted algebra is then:
%\begin{align}
%    \tilde{P} &= P+\xi R_{12}^+& \tilde{K}&=K-\xi^{-1}R^{12}_- &\tilde{D}&=D-\frac12(R_1^1+R_2^2)
%\end{align}
%For the four point function, the topological condition on the OPE of this correlator is then
%\begin{align}
%    &\lim_{\zeta\rightarrow \chi} \left(f^{(1,0,0)}+2\bar{\zeta} f^{(0,1,0)}-2(1-\bar{\zeta}) f^{(0,1,0)}\right)=0
%\end{align}
%which can be written as
%\begin{align}
%    (\frac 12 \partial_\chi +\partial_{\zeta})f(\chi,\zeta \bar{\zeta},(1-\zeta)(1-\bar{\zeta}))|_{\zeta \rightarrow \chi}=0
%\end{align}
\section{$su(N)$}
Following the classification of \cite{Agmon:2020pde}, the requirements are met by the representations of type $A_1\bar{A}_1$ for the supergroup $su(1,1|\frac 12 \mc{N})$. In the case of extended supersymmetry, there are some choices to be made in terms of the topological charges used. 
\begin{itemize}
\item {$\mc{N}=2$}\par 
For $\mc{N}=2$, there are no R-symmetry indices and correspondingly not enough supercharges to form a topological operator. \par 
\item {$\mc{N}=4$}\par 
For $\mc{N}=4$, we have the operators of type $A_1\bar{A}_1$ annihilated by 
\begin{align}
    Q_1&&\bar{Q}^2
\end{align}
with the unitarity bound 
\begin{align}
    \Delta=\frac 12 R
\end{align}
These are precisely the conditions specified above, so we have the Ward identity on the $n$-point function
\begin{align}
    \langle[0]^{(R)}_{\frac R 2}(t_1,y_1)....[0]^{(R)}_{\frac R 2}(t_n,y_n) \rangle &=A_{n,R,\frac R 2} (t_i,y_i) f_{n,A_1\bar{A}_1}(\chi_i,\rho_i)
\end{align}
\begin{align}
    \left(\partial_{\chi_j}+\partial_{\rho_j}\right) f_{n,A_1\bar{A}_1}(\chi_i,\rho_i)&=0
\end{align}
\item {$\mc{N}=6$}
Let us now consider the Ward identities constraining correlators of operators charged under an $su(2)$ subgroup of the R-symmetry
\begin{align}
    \Phi_{\Delta,(R_1,0,0...)}&&\bar{\Phi}_{\Delta,(0,...,0,R_{N-1})}
\end{align}
are annihilated by the supercharges
\begin{align}
    \{Q_1,\bar{Q}^N\}
\end{align}
We can form the topological charges:
\begin{align}
    \mathbb{Q} = Q_1+S_N&&\mathbb{Q}^\dag = \bar{Q}^N +\bar{S}^1
\end{align}
that generate the twisted conformal algebra
\begin{align}
    \tilde{P} &=P+E_- &\tilde{K}&= P-E_-&\tilde{D}&=D-
\end{align}
such that we recover the $su(2)$ Ward identities above for the field
\begin{align}
        \Phi_{\Delta,(\lambda_1,0,...)}(t,y)&=\text{e}^{-t P-yE_-}\Phi_{\Delta,(\lambda_1,0,...)}(0,0)
\end{align}
where:
\begin{align}
    E_-&=R^1_N&E_+&=R^{N}_1&H_0 = \frac{1}{2}(R_1^1-R_N^N)=\frac 12 \sum_{i=1}^{N-1}\lambda_i
\end{align}
Since the differential operators can only act on one point each, we can separatly define 
\begin{align}
        \bar{\Phi}_{\Delta,(0,...,0,\lambda)}(t,\bar{y})&=\text{e}^{-t P-\bar{y}E_-}\Phi_{\Delta,(0,...0,,\lambda)}(0,0)
\end{align}
So that the $u(1)$-invariant $su(2)$ distance is naturally defined:
\begin{align}
 (1\bar{2})&=y_1-\bar{y}_2
\end{align}
This gives the same Ward identities as before:
\begin{align}
    \langle \Phi_{\Delta,(\Delta,0,...)}\bar{\Phi}_{\Delta,(0,...,0,\Delta)}\Phi_{\Delta,(\Delta,0,...)}\bar{\Phi}_{\Delta,(0,...,0,\Delta)}\rangle &=\left(\frac{(1\bar{2})(3\bar{4})}{t_{12}t_{34}}\right)^\Delta f(\chi_i,\rho_i)\\
    \left(\partial_{\chi_i}+\partial_{\rho_i}\right)f(\chi_i,\rho_i)|_{\rho_i\rightarrow \chi_i}&=0
\end{align}
where we use the interdependence of the R-symmetry variables to find u(1)-invariant terms, for example for the 4-point function:
\begin{align}
    \rho &= \frac{(1\bar{2})(3\bar{4})}{(13)(\bar{2}\bar{4})}&1-\rho&=-\frac{(1\bar{4})(3\bar{2})}{(13)(\bar{2}\bar{4})}
\end{align}
So that:
\begin{align}
   \frac{\rho}{\rho-1} = \frac{(1\bar{2})(3\bar{4})}{(1\bar{4})(3\bar{2})}
\end{align}
Equivalently, one can find an embedding complex vector $Y_a,\bar{Y}^a$ as in \cite{Dolan:2004iy} such that 
\begin{align}
    Y_i\cdot \bar{Y}_j= y_{ij}.
\end{align}
In this setting, it is possible to express $\{E_-, E_+,H_0\}$ as differential operators acting only on $y$ such that the $su(2)$ results hold.
\end{itemize}
\bibliography{bib}

%bibliography generated by nb.bst v1.01 (C) 2003-2010 Niklas Beisert
\begin{thebibliography}{10}
\ifx\href\asklfhas\newcommand{\href}[2]{#2}\fi
\ifx\arxivref\asklfhas\newcommand{\arxivref}[2]{\href{http://arxiv.org/abs/#1}{#2}}\fi
\ifx\doiref\asklfhas\newcommand{\doiref}[2]{\href{http://dx.doi.org/#1}{#2}}\fi
\raggedright
\small
\parskip 0pt

\bibitem{Billo:2016cpy}
M.~Bill\`o, V.~Gon\c{c}alves, E.~Lauria and M.~Meineri,
\textit{``{Defects in conformal field theory}''},
\textsf{\doiref{10.1007/JHEP04(2016)091}{JHEP~1604,~091~(2016)}},
\texttt{\arxivref{1601.02883}{arxiv:1601.02883}}.

\bibitem{Giombi:2018qox}
S.~Giombi and S.~Komatsu,
\textit{``{Exact Correlators on the Wilson Loop in $\mathcal{N}=4$ SYM:
  Localization, Defect CFT, and Integrability}''},
\textsf{\doiref{10.1007/JHEP05(2018)109}{JHEP~1805,~109~(2018)}},
\texttt{\arxivref{1802.05201}{arxiv:1802.05201}},
[Erratum: JHEP 11, 123 (2018)].

\bibitem{Dolan:2004mu}
F.~A.~Dolan, L.~Gallot and E.~Sokatchev,
\textit{``{On four-point functions of 1/2-BPS operators in general
  dimensions}''},
\textsf{\doiref{10.1088/1126-6708/2004/09/056}{JHEP~0409,~056~(2004)}},
\texttt{\arxivref{hep-th/0405180}{hep-th/0405180}}.

\bibitem{Eden:2000qp}
B.~U.~Eden, P.~S.~Howe, A.~Pickering, E.~Sokatchev and P.~C.~West,
\textit{``{Four point functions in N=2 superconformal field theories}''},
\textsf{\doiref{10.1016/S0550-3213(00)00218-2}{Nucl.~Phys.~B~581,~523~(2000)}},
\texttt{\arxivref{hep-th/0001138}{hep-th/0001138}}.

\bibitem{Nirschl:2004pa}
M.~Nirschl and H.~Osborn,
\textit{``{Superconformal Ward identities and their solution}''},
\textsf{\doiref{10.1016/j.nuclphysb.2005.01.013}{Nucl.~Phys.~B~711,~409~(2005)}},
\texttt{\arxivref{hep-th/0407060}{hep-th/0407060}}.

\bibitem{Hartwell:1994rp}
G.~G.~Hartwell and P.~S.~Howe,
\textit{``{(N, p, q) harmonic superspace}''},
\textsf{\doiref{10.1142/S0217751X95001820}{Int.~J.~Mod.~Phys.~A~10,~3901~(1995)}},
\texttt{\arxivref{hep-th/9412147}{hep-th/9412147}}.

\bibitem{Dolan:2004iy}
F.~A.~Dolan and H.~Osborn,
\textit{``{Conformal partial wave expansions for N=4 chiral four point
  functions}''},
\textsf{\doiref{10.1016/j.aop.2005.07.005}{Annals~Phys.~321,~581~(2006)}},
\texttt{\arxivref{hep-th/0412335}{hep-th/0412335}}.

\bibitem{Beem:2013qxa}
C.~Beem, L.~Rastelli and B.~C.~van~Rees,
\textit{``{The $\mathcal N=4$ Superconformal Bootstrap}''},
\textsf{\doiref{10.1103/PhysRevLett.111.071601}{Phys.~Rev.~Lett.~111,~071601~(2013)}},
\texttt{\arxivref{1304.1803}{arxiv:1304.1803}}.

\bibitem{Alday:2014qfa}
L.~F.~Alday and A.~Bissi,
\textit{``{Generalized bootstrap equations for $ \mathcal{N}=4 $ SCFT}''},
\textsf{\doiref{10.1007/JHEP02(2015)101}{JHEP~1502,~101~(2015)}},
\texttt{\arxivref{1404.5864}{arxiv:1404.5864}}.

\bibitem{Chester:2014fya}
S.~M.~Chester, J.~Lee, S.~S.~Pufu and R.~Yacoby,
\textit{``{The $ \mathcal{N}=8 $ superconformal bootstrap in three
  dimensions}''},
\textsf{\doiref{10.1007/JHEP09(2014)143}{JHEP~1409,~143~(2014)}},
\texttt{\arxivref{1406.4814}{arxiv:1406.4814}}.

\bibitem{Beem:2014zpa}
C.~Beem, M.~Lemos, P.~Liendo, L.~Rastelli and B.~C.~van~Rees,
\textit{``{The $ \mathcal{N}=2 $ superconformal bootstrap}''},
\textsf{\doiref{10.1007/JHEP03(2016)183}{JHEP~1603,~183~(2016)}},
\texttt{\arxivref{1412.7541}{arxiv:1412.7541}}.

\bibitem{Belitsky:2014zha}
A.~V.~Belitsky, S.~Hohenegger, G.~P.~Korchemsky and E.~Sokatchev,
\textit{``{N=4 superconformal Ward identities for correlation functions}''},
\textsf{\doiref{10.1016/j.nuclphysb.2016.01.008}{Nucl.~Phys.~B~904,~176~(2016)}},
\texttt{\arxivref{1409.2502}{arxiv:1409.2502}}.

\bibitem{Giombi:2023zte}
S.~Giombi, S.~Komatsu, B.~Offertaler and J.~Shan,
\textit{``{Boundary reparametrizations and six-point functions on the AdS$_2$
  string}''},
\texttt{\arxivref{2308.10775}{arxiv:2308.10775}}.

\bibitem{Antunes:2023dlk}
A.~Antunes, S.~Harris, A.~Kaviraj and V.~Schomerus,
\textit{``{Lining up a Positive Semi-Definite Six-Point Bootstrap}''},
\texttt{\arxivref{2312.11660}{arxiv:2312.11660}}.

\bibitem{Freedman:1998tz}
D.~Z.~Freedman, S.~D.~Mathur, A.~Matusis and L.~Rastelli,
\textit{``{Correlation functions in the CFT(d) / AdS(d+1) correspondence}''},
\textsf{\doiref{10.1016/S0550-3213(99)00053-X}{Nucl.~Phys.~B~546,~96~(1999)}},
\texttt{\arxivref{hep-th/9804058}{hep-th/9804058}}.

\bibitem{Rastelli:2017udc}
L.~Rastelli and X.~Zhou,
\textit{``{How to Succeed at Holographic Correlators Without Really Trying}''},
\textsf{\doiref{10.1007/JHEP04(2018)014}{JHEP~1804,~014~(2018)}},
\texttt{\arxivref{1710.05923}{arxiv:1710.05923}}.

\bibitem{Bliard:2022xsm}
G.~Bliard,
\textit{``{Notes on n-point Witten diagrams in AdS$_{2}$}''},
\textsf{\doiref{10.1088/1751-8121/ac7f6b}{J.~Phys.~A~55,~325401~(2022)}},
\texttt{\arxivref{2204.01659}{arxiv:2204.01659}}.

\bibitem{Paulos:2011ie}
M.~F.~Paulos,
\textit{``{Towards Feynman rules for Mellin amplitudes}''},
\textsf{\doiref{10.1007/JHEP10(2011)074}{JHEP~1110,~074~(2011)}},
\texttt{\arxivref{1107.1504}{arxiv:1107.1504}}.

\bibitem{Mack:2009mi}
G.~Mack,
\textit{``{D-independent representation of Conformal Field Theories in D
  dimensions via transformation to auxiliary Dual Resonance Models. Scalar
  amplitudes}''},
\texttt{\arxivref{0907.2407}{arxiv:0907.2407}}.

\bibitem{Penedones:2010ue}
J.~Penedones,
\textit{``{Writing CFT correlation functions as AdS scattering amplitudes}''},
\textsf{\doiref{10.1007/JHEP03(2011)025}{JHEP~1103,~025~(2011)}},
\texttt{\arxivref{1011.1485}{arxiv:1011.1485}}.

\bibitem{Liendo:2016ymz}
P.~Liendo and C.~Meneghelli,
\textit{``{Bootstrap equations for $ \mathcal{N} $ = 4 SYM with defects}''},
\textsf{\doiref{10.1007/JHEP01(2017)122}{JHEP~1701,~122~(2017)}},
\texttt{\arxivref{1608.05126}{arxiv:1608.05126}}.

\bibitem{Liendo:2018ukf}
P.~Liendo, C.~Meneghelli and V.~Mitev,
\textit{``{Bootstrapping the half-BPS line defect}''},
\textsf{\doiref{10.1007/JHEP10(2018)077}{JHEP~1810,~077~(2018)}},
\texttt{\arxivref{1806.01862}{arxiv:1806.01862}}.

\bibitem{Cavaglia:2022qpg}
A.~Cavagli\`a, N.~Gromov, J.~Julius and M.~Preti,
\textit{``{Bootstrability in defect CFT: integrated correlators and sharper
  bounds}''},
\textsf{\doiref{10.1007/JHEP05(2022)164}{JHEP~2205,~164~(2022)}},
\texttt{\arxivref{2203.09556}{arxiv:2203.09556}}.

\bibitem{Cavaglia:2021bnz}
A.~Cavagli\`a, N.~Gromov, J.~Julius and M.~Preti,
\textit{``{Integrability and conformal bootstrap: One dimensional defect
  conformal field theory}''},
\textsf{\doiref{10.1103/PhysRevD.105.L021902}{Phys.~Rev.~D~105,~L021902~(2022)}},
\texttt{\arxivref{2107.08510}{arxiv:2107.08510}}.

\bibitem{Ferrero:2021bsb}
P.~Ferrero and C.~Meneghelli,
\textit{``{Bootstrapping the half-BPS line defect CFT in N=4 supersymmetric
  Yang-Mills theory at strong coupling}''},
\textsf{\doiref{10.1103/PhysRevD.104.L081703}{Phys.~Rev.~D~104,~L081703~(2021)}},
\texttt{\arxivref{2103.10440}{arxiv:2103.10440}}.

\bibitem{Barrat:2024nod}
J.~Barrat,
\textit{``{Line defects in conformal field theory}''},
\texttt{\arxivref{2401.10336}{arxiv:2401.10336}}.

\bibitem{Bliard:2023zpe}
G.~J.~S.~Bliard,
\textit{``{Perturbative and non-perturbative analysis of defect correlators in
  AdS/CFT}''},
\texttt{\arxivref{2310.18137}{arxiv:2310.18137}}.

\bibitem{Ferrero:2023gnu}
P.~Ferrero and C.~Meneghelli,
\textit{``{Unmixing the Wilson line defect CFT. Part II: analytic
  bootstrap}''},
\texttt{\arxivref{2312.12551}{arxiv:2312.12551}}.

\bibitem{Ferrero:2023znz}
P.~Ferrero and C.~Meneghelli,
\textit{``{Unmixing the Wilson line defect CFT. Part I. Spectrum and
  kinematics}''},
\textsf{\doiref{10.1007/JHEP05(2024)090}{JHEP~2405,~090~(2024)}},
\texttt{\arxivref{2312.12550}{arxiv:2312.12550}}.

\bibitem{Cooke:2017qgm}
M.~Cooke, A.~Dekel and N.~Drukker,
\textit{``{The Wilson loop CFT: Insertion dimensions and structure constants
  from wavy lines}''},
\textsf{\doiref{10.1088/1751-8121/aa7db4}{J.~Phys.~A~50,~335401~(2017)}},
\texttt{\arxivref{1703.03812}{arxiv:1703.03812}}.

\bibitem{Drukker:2006xg}
N.~Drukker and S.~Kawamoto,
\textit{``{Small deformations of supersymmetric Wilson loops and open
  spin-chains}''},
\textsf{\doiref{10.1088/1126-6708/2006/07/024}{JHEP~0607,~024~(2006)}},
\texttt{\arxivref{hep-th/0604124}{hep-th/0604124}}.

\bibitem{Correa:2012at}
D.~Correa, J.~Henn, J.~Maldacena and A.~Sever,
\textit{``{An exact formula for the radiation of a moving quark in N=4 super
  Yang Mills}''},
\textsf{\doiref{10.1007/JHEP06(2012)048}{JHEP~1206,~048~(2012)}},
\texttt{\arxivref{1202.4455}{arxiv:1202.4455}}.

\bibitem{Gaberdiel:2008fn}
M.~R.~Gaberdiel, A.~Konechny and C.~Schmidt-Colinet,
\textit{``{Conformal perturbation theory beyond the leading order}''},
\textsf{\doiref{10.1088/1751-8113/42/10/105402}{J.~Phys.~A~42,~105402~(2009)}},
\texttt{\arxivref{0811.3149}{arxiv:0811.3149}}.

\bibitem{Correa:2012nk}
D.~Correa, J.~Henn, J.~Maldacena and A.~Sever,
\textit{``{The cusp anomalous dimension at three loops and beyond}''},
\textsf{\doiref{10.1007/JHEP05(2012)098}{JHEP~1205,~098~(2012)}},
\texttt{\arxivref{1203.1019}{arxiv:1203.1019}}.

\bibitem{Drukker:2009sf}
N.~Drukker and J.~Plefka,
\textit{``{Superprotected n-point correlation functions of local operators in
  N=4 super Yang-Mills}''},
\textsf{\doiref{10.1088/1126-6708/2009/04/052}{JHEP~0904,~052~(2009)}},
\texttt{\arxivref{0901.3653}{arxiv:0901.3653}}.

\bibitem{Barrat:2021tpn}
J.~Barrat, P.~Liendo, G.~Peveri and J.~Plefka,
\textit{``{Multipoint correlators on the supersymmetric Wilson line defect
  CFT}''},
\textsf{\doiref{10.1007/JHEP08(2022)067}{JHEP~2208,~067~(2022)}},
\texttt{\arxivref{2112.10780}{arxiv:2112.10780}}.

\bibitem{Baume:2019aid}
F.~Baume, M.~Fuchs and C.~Lawrie,
\textit{``{Superconformal Blocks for Mixed 1/2-BPS Correlators with $SU(2)$
  R-symmetry}''},
\textsf{\doiref{10.1007/JHEP11(2019)164}{JHEP~1911,~164~(2019)}},
\texttt{\arxivref{1908.02768}{arxiv:1908.02768}}.

\bibitem{Cornagliotto:2017dup}
M.~Cornagliotto, M.~Lemos and V.~Schomerus,
\textit{``{Long Multiplet Bootstrap}''},
\textsf{\doiref{10.1007/JHEP10(2017)119}{JHEP~1710,~119~(2017)}},
\texttt{\arxivref{1702.05101}{arxiv:1702.05101}}.

\bibitem{Buric:2020zlb}
I.~Buric,
\textit{``{Harmonic Analysis in d-dimensional Superconformal Field Theory}''},
\textsf{\doiref{10.3842/SIGMA.2021.007}{SIGMA~17,~007~(2021)}},
\texttt{\arxivref{2009.00393}{arxiv:2009.00393}}.

\bibitem{Bissi:2015qoa}
A.~Bissi and T.~\L{}ukowski,
\textit{``{Revisiting $ \mathcal{N}=4 $ superconformal blocks}''},
\textsf{\doiref{10.1007/JHEP02(2016)115}{JHEP~1602,~115~(2016)}},
\texttt{\arxivref{1508.02391}{arxiv:1508.02391}}.

\bibitem{Buric:2019rms}
I.~Buric, V.~Schomerus and E.~Sobko,
\textit{``{Superconformal Blocks: General Theory}''},
\textsf{\doiref{10.1007/JHEP01(2020)159}{JHEP~2001,~159~(2020)}},
\texttt{\arxivref{1904.04852}{arxiv:1904.04852}}.

\bibitem{Gimenez-Grau:2019hez}
A.~Gimenez-Grau and P.~Liendo,
\textit{``{Bootstrapping line defects in $\mathcal{N}=2$ theories}''},
\textsf{\doiref{10.1007/JHEP03(2020)121}{JHEP~2003,~121~(2020)}},
\texttt{\arxivref{1907.04345}{arxiv:1907.04345}}.

\bibitem{Bianchi:2020hsz}
L.~Bianchi, G.~Bliard, V.~Forini, L.~Griguolo and D.~Seminara,
\textit{``{Analytic bootstrap and Witten diagrams for the ABJM Wilson line as
  defect CFT$_{1}$}''},
\textsf{\doiref{10.1007/JHEP08(2020)143}{JHEP~2008,~143~(2020)}},
\texttt{\arxivref{2004.07849}{arxiv:2004.07849}}.

\bibitem{Barrat:2021yvp}
J.~Barrat, A.~Gimenez-Grau and P.~Liendo,
\textit{``{Bootstrapping holographic defect correlators in $ \mathcal{N} $ = 4
  super Yang-Mills}''},
\textsf{\doiref{10.1007/JHEP04(2022)093}{JHEP~2204,~093~(2022)}},
\texttt{\arxivref{2108.13432}{arxiv:2108.13432}}.

\bibitem{Drukker:2008pi}
N.~Drukker and J.~Plefka,
\textit{``{The Structure of n-point functions of chiral primary operators in
  N=4 super Yang-Mills at one-loop}''},
\textsf{\doiref{10.1088/1126-6708/2009/04/001}{JHEP~0904,~001~(2009)}},
\texttt{\arxivref{0812.3341}{arxiv:0812.3341}}.

\bibitem{Meneghelli:2022gps}
C.~Meneghelli and M.~Tr\'epanier,
\textit{``{Bootstrapping string dynamics in the 6d \ensuremath{\mathscr{N}} =
  (2, 0) theories}''},
\textsf{\doiref{10.1007/JHEP07(2023)165}{JHEP~2307,~165~(2023)}},
\texttt{\arxivref{2212.05020}{arxiv:2212.05020}}.

\bibitem{Dolan:2000ut}
F.~A.~Dolan and H.~Osborn,
\textit{``{Conformal four point functions and the operator product
  expansion}''},
\textsf{\doiref{10.1016/S0550-3213(01)00013-X}{Nucl.~Phys.~B~599,~459~(2001)}},
\texttt{\arxivref{hep-th/0011040}{hep-th/0011040}}.

\bibitem{Chester:2014mea}
S.~M.~Chester, J.~Lee, S.~S.~Pufu and R.~Yacoby,
\textit{``{Exact Correlators of BPS Operators from the 3d Superconformal
  Bootstrap}''},
\textsf{\doiref{10.1007/JHEP03(2015)130}{JHEP~1503,~130~(2015)}},
\texttt{\arxivref{1412.0334}{arxiv:1412.0334}}.

\bibitem{Liendo:2015cgi}
P.~Liendo, C.~Meneghelli and V.~Mitev,
\textit{``{On Correlation Functions of BPS Operators in 3d ${\mathcal{N}}$ = 6
  Superconformal Theories}''},
\textsf{\doiref{10.1007/s00220-016-2715-7}{Commun.~Math.~Phys.~350,~387~(2017)}},
\texttt{\arxivref{1512.06072}{arxiv:1512.06072}}.

\bibitem{Peveri:2023qip}
G.~Peveri,
\textit{``{Correlators on the Wilson Line Defect CFT}''},
\texttt{\arxivref{2310.17358}{arxiv:2310.17358}}.

\bibitem{Correa:2021sky}
D.~H.~Correa, V.~I.~Giraldo-Rivera and M.~Lagares,
\textit{``{On the abundance of supersymmetric strings in AdS$_{3}$ \texttimes{}
  S $^{3}$ \texttimes{} S $^{3}$ \texttimes{} S $^{1}$ describing BPS line
  operators}''},
\textsf{\doiref{10.1088/1751-8121/ac354d}{J.~Phys.~A~54,~505401~(2021)}},
\texttt{\arxivref{2108.09380}{arxiv:2108.09380}}.

\bibitem{Lemos:2015awa}
M.~Lemos and P.~Liendo,
\textit{``{Bootstrapping $ \mathcal{N}=2 $ chiral correlators}''},
\textsf{\doiref{10.1007/JHEP01(2016)025}{JHEP~1601,~025~(2016)}},
\texttt{\arxivref{1510.03866}{arxiv:1510.03866}}.

\bibitem{Beem:2016wfs}
C.~Beem, L.~Rastelli and B.~C.~van~Rees,
\textit{``{More ${\mathcal N}=4$ superconformal bootstrap}''},
\textsf{\doiref{10.1103/PhysRevD.96.046014}{Phys.~Rev.~D~96,~046014~(2017)}},
\texttt{\arxivref{1612.02363}{arxiv:1612.02363}}.

\bibitem{multipoints:sophie}
J.~Barrat, C.~Meneghelli and S.~Mueller,
\textit{``to appear''}.

\bibitem{Agmon:2020pde}
N.~B.~Agmon and Y.~Wang,
\textit{``{Classifying Superconformal Defects in Diverse Dimensions Part I:
  Superconformal Lines}''},
\texttt{\arxivref{2009.06650}{arxiv:2009.06650}}.

\bibitem{Guerrini:2023rdw}
L.~Guerrini,
\textit{``{On protected defect correlators in 3d $ \mathcal{N}
  $\ensuremath{\geq} 4 theories}''},
\textsf{\doiref{10.1007/JHEP10(2023)100}{JHEP~2310,~100~(2023)}},
\texttt{\arxivref{2301.07035}{arxiv:2301.07035}}.

\end{thebibliography}
\bibliographystyle{nb}
	
\end{document}